\documentclass[twocolumn,showpacs,showkeys,amsmath,amssymb,superscriptaddress,floatfix,nofootinbib]{revtex4-1}

\usepackage{graphicx}
\usepackage{bm} 
\usepackage{subfigure}
	
\def\vec#1{\mathchoice{\mbox{\boldmath$\displaystyle#1$}}
{\mbox{\boldmath$\textstyle#1$}}
{\mbox{\boldmath$\scriptstyle#1$}}
{\mbox{\boldmath$\scriptscriptstyle#1$}}}
\makeatletter

\begin{document}
 
\title{Signatures of collective flow in high multiplicity pp
  collisions. 
}

\author{Adam Kisiel} 
\email{Adam.Kisiel@cern.ch}
\affiliation{Physics Department, CERN, CH-1211, Gen\`eve 23, Switzerland}
   
\begin{abstract}
A blast-wave parametrization, including a full set of hadronic
resonances, is used to model a small system, with total particle
multiplicity comparable to the one measured in the high-multiplicity
pp collisions at the LHC. Calculations are preformed for three
cases: with negligible, regular and strong radial flow on the
blast-wave hypersurface. We investigate the effects of flow on
inclusive $p_{T}$ spectra as well as on 1D and 3D femtoscopic radii
for pions. Special emphasis is put on the role of pions from resonance
decays. In particular we show that they magnify the flow effects
present in the blast-wave stage and significantly influence the shape
of the correlation functions. A specific observable, the
$R^{E}_{out}/R^{G}_{side}$ ratio is proposed as a sensitive probe of
the   collective effects. Model results for the high multiplicity pp
collisions, for scenarios with small and large radial flow are
compared.     
\end{abstract}



\maketitle 


\section{Introduction}
\label{sec:intro}

The pp program at the LHC has successfully started, and has
already produced a large set of minimum-bias measurements from the
experiments~\cite{Aamodt:2009dt,Aamodt:2010ft,Aamodt:2010pp,Aamodt:2010jj,Aamodt:2010my,Khachatryan:2010xs,Khachatryan:2010un,Khachatryan:2010us}.
Of particular interest is the fact that the multiplicity range
measured at $\sqrt{s} = 7$~TeV overlaps with the one measured in
peripheral CuCu collisions at ultra-relativistic energies of
$\sqrt{s_{NN}} = 62$ and $200$~GeV at RHIC. This allows for comparison
of observables between p+p and heavy-ion collisions at the same
measured final state multiplicity $dN_{ch}/d\eta$. In heavy-ion
collisions, $dN_{ch}/d\eta$ is the scaling variable for many soft
physics observables, including the inclusive particle $p_{\rm T}$ spectra, pion
femtoscopic radii~\cite{Lisa:2005dd}, strange particle yields etc. It
is therefore interesting whether a high-multiplicity pp collision
resembles a heavy-ion one.   

This question has become even more relevant with the recent
observation of the long-range $\eta$ ridge in high multiplicity pp
collisions by the CMS experiment~\cite{Collaboration:2010gv}. There
are attempts to explain the phenomenon in the Color Glass Condensate
framework~\cite{Dumitru:2010iy} or as an elliptic 
flow~\cite{Bozek:2010pb}. Both of these critically depend on the
observation of radial flow (or space-momentum correlation) in such
collisions. A recent study with EPOS model~\cite{Werner:2010ny}
suggests that such flow has actually been seen in the ALICE pion
femtoscopy data~\cite{Aamodt:2010jj}. Similar topics have been
investigated in the framework of the rescattering
model~\cite{Humanic:2006ib}. A comparison of pion femtoscopic radii in
pp and heavy-ion collisions at lower energy $\sqrt{s} = 200$~GeV have
been carried out by the STAR
experiment~\cite{Aggarwal:2010bw}. Scaling between radii in pp and
heavy-ion collisions has been observed vs. the pair momentum,
suggesting that there might be a common physics mechanism, i.e. radial
flow, driving them in both cases. 

The hydrodynamic scenario is well applicable to heavy-ion collisions
at RHIC. It produces specific space-momentum correlation patterns,
which are commonly referred to as flow. The system created 
in the heavy-ion collision expands rapidly outwards, showing a 
strong radial flow, which is observed in the modification of the
single particle inclusive $p_{\rm T}$ spectra shape. The spectrum
is an observable depending only on the momenta of the particles,  so
its connection to space--time can only be indirectly inferred. To
directly access space--time information we employ femtoscopic
techniques. The fall of the ``femtoscopic radii'' (sometimes called
``HBT radii'') with particle's $m_T$ can be interpreted as the
decrease of ``lengths of homogeneity'', a direct  consequence of
radial and longitudinal flow~\cite{Akkelin:1995gh}. The simultaneous 
description of the $p_{\rm T}$ spectra and the femtoscopic radii dependence
on $m_{T}$ in heavy-ion collisions can be achieved with the
``blast-wave''
parametrization~\cite{Schnedermann:1993ws,Retiere:2003kf,Kisiel:2006is}.
In this work we test what results such an approach brings when applied
to small systems. We characterize the size of the system by the
average charged particle multiplicity per unit of  pseudorapidity that it
produces. This means that simple ``blast-wave'' models cannot be used,
as abundances of each particle species are free parameters there. In
contrast the {\tt THERMINATOR} model~\cite{Kisiel:2005hn} has both
necessary features: it can use the traditional ``blast-wave''
hypersurface and provides absolute particle multiplicities. 

The usage of the {\rm THERMINATOR}  makes it possible to test
the impact of the strongly decaying resonances. As it was shown
in~\cite{Kisiel:2006is}, their introduction increases the apparent
size of the system produced in heavy-ion collisions  by approximately
$1$~fm and leads to the development of long-range tails in the emission
function of pions. These effects were important but not dominant in
the AuAu system (with the average size of $6$~fm), but they may
dominate in a pp collision where we expect the initial system to have
a size of the order of $1$~fm.  

The paper is organized as follows. In Section~\ref{sec:blastwave} we
briefly introduce the {\tt THERMINATOR} model and its blast-wave
variant. In Section~\ref{sec:simulations} we describe the choice of
parameters for the three simulations that were performed. In
Section~\ref{sec:results} we describe the results for the inclusive
transverse momentum $p_{\rm T}$ spectra and the 1-dimensional femtoscopic
simulations and explain the features observed there. In
Section~\ref{sec:analysis3d} we show the results of the 3D femtoscopic
analysis and propose the most robust and sensitive observables that
can be used to search for collectivity in the pp collisions at large
multiplicities, e.g. at the LHC. In Section~\ref{sec:origins} we
discuss the origin of the effects seen in previous Sections.
 
\section{{\tt THERMINATOR} model and femtoscopic formalism}
\label{sec:blastwave}

In this work we use the {\tt THERMINATOR} model~\cite{Kisiel:2005hn},
more specifically its variant that uses the blast-wave freeze-out
hypersurface with the possibility to adjust the space-time correlation
at the freeze-out hypersurface. For a detailed description of the
model, its full theoretical basis and formalism
see~\cite{Kisiel:2006is}. In short, average per-event particle
abundances are calculated from the chemical model, where the
temperature $T$ and bariochemical potential $\mu_B$ are model
parameters, and isospin $\mu_I$ and strange $\mu_S$ potential are
calculated from conservation laws. For each event the number of
particles of each type (we use 381 particle types taken from
PDG~\cite{Yao:2006px}) is randomly generated from a Poissonian 
distribution. Then each particle is put on the freeze-out hypersurface
according to the generalized blast-wave prescription~\cite{Kisiel:2006is}:
\begin{widetext}
\begin{eqnarray}
{dN \over dy d\varphi p_\perp dp_\perp d\alpha_\parallel  d\phi \rho  d\rho \, } 
&=& {1 \over (2\pi)^3} (\tau + a\rho)
\left[ m_\perp \hbox{cosh}(\alpha_\parallel-y) 
- a \, p_\perp \cos(\phi-\varphi)\right] \nonumber \\
& & \times \left\{
\exp\left[ { \beta m_\perp  \over \sqrt{1 - v_\perp^2}} \,\,
\hbox{cosh}(\alpha_\parallel-y) - { \beta p_\perp   v_\perp \over \sqrt{1 - v_\perp^2}}
 \cos(\phi-\varphi) - \beta \mu \right] 
\pm 1 \right\}^{-1} ,
\label{modA2}
\end{eqnarray}
\end{widetext}
where $a$, $\tau$, $\rho_{\rm max}$, are parameters of the
model. $\rho_{\rm max}$ is the transverse size of the system, $\tau$
is the proper time of emission, and $a$ determines the transverse
position ($\rho$) vs. time ($t$) correlation at the hypersurface. Of
particular importance is the form of the radial flow velocity, which
depends on particle emission position:  
\begin{equation}
v_{\perp} = \frac {\rho/\rho_{\rm max}} {v_{\rm T} + \rho/\rho_{\rm
    max}} ,
\label{eq:vprof}
\end{equation}
where $v_{\rm T}$ is a parameter which controls the amount of flow in the
system. Large values of $v_{\rm T}$ give small flow, small values of $v_{\rm T}$
give large flow, growing semi-linearly from $0$ at the center of
the source to the maximum value of $1/(1+v_{\rm T})$ at the edge. $y$,
$p_{\rm T}$ and $\varphi$ are the components of the momentum of the
particle (rapidity, transverse momentum and azimuthal angle
respectively), while $\alpha_\parallel$, $\rho$ and $\phi$ are the
corresponding coordinates of the emission point. As the last step, all
unstable particles are propagated and decayed, in cascades if
necessary, until only stable particles remain. The emission point of
the particle is either its creation point on the hypersurface (we call
such particles ``primordial'') or a decay point of its parent
resonance.  

Final state particle rescatterings after the emission are not
included. We have estimated that for the particle densities in this
work, each particle would rescatter on the average 0.8
times. Most of the rescattering points would be very close to the
original emission point, so the inclusion of rescattering would not
change the emission shapes significantly.  

Further in the work we show results of the analysis of the identical
pion femtoscopic correlations (sometimes called ``Bose-Einstein
correlations'' or ``HBT correlations''). All the analyzes were
performed using the standard ``two-particle'' method: pions from the
same event were combined into pairs, and assigned a weight equal to
their wave-function squared, to form the signal. The pair wave
function contained only the relevant symmetrization; final state
coulomb and strong interactions were ignored as for pions they are
small and are not important for the determination of the femtoscopic
radii. Then, pions from different events were combined into pairs to
form the background. The correlation function was constructed by
dividing the signal by the background. It was then fitted with various
functional forms, which will be described in detail in
Section~\ref{sec:results}. For an extensive description of the
two-particle method and the formalism see~\cite{Kisiel:2006is}. 

\section{Simulation description}
\label{sec:simulations}

\begin{figure}[tb]
\begin{center}
\includegraphics[angle=0,width=0.4 \textwidth]{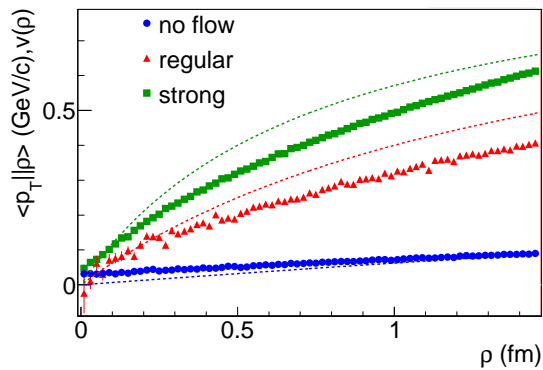}
\end{center}
\vspace{-6.5mm}
\caption{(Color on-line) Average common transverse flow of
  primordial pions (see Eq.\eqref{eq:ptparrt}) as a function of the
  distance to the central axis of the source $\rho$. ``No flow'' results are
  shown as blue circles, ``regular flow'' as red triangles, ``strong
  flow'' as green squares. The input model flow velocity profiles (see
  Eq.~\eqref{eq:vprof}) are shown as dashed lines. 
\label{fig:vprofiles}}
\end{figure}

We performed the simulations with the blast-wave model with the
following parameters: transverse size of the freeze-out hypersurface
$\rho_{max} = 1.5$~fm, proper time of freeze-out $\tau_{0} =
2.0$~fm/c, freeze-out temperature $T_{f} = 165.6$~MeV, bariochemical
potential $\mu_B = 28.5$~MeV. The freeze-out hypersurface was slightly
sloped in the $\rho-t$ plane ($a = - 0.2$). The three simulations
differed only in the value of the $v_{\rm T}$ parameter; the ``no-flow''
had $v_{\rm T} = 10$, the ``regular flow'': $v_{\rm T} = 1$ and the ``strong
flow'': $v_{\rm T} = 0.5$. The resulting flow velocity profiles can be
seen in Fig.~\ref{fig:vprofiles} as dashed lines. The common radial
flow present in the model results in strong space-momentum $\rho-p_{\rm T}$
correlations of the produced particles, which is reflected in the the
common radial flow: 
\begin{equation}
\left <p_{\rm T}\parallel \rho \right> = \left < \frac {p_{\rm T} \rho
    \cos(\phi-\varphi)} {\rho} \right >  ,
\label{eq:ptparrt}
\end{equation}
where averaging is done over particles. The results are shown in
Fig.~\ref{fig:vprofiles} as symbols. The plot illustrates that the
``no flow'' scenario has very little $\rho-p_{\rm T}$ correlations, while in
the ``strong flow'' scenario this correlation is significant. For each
of the scenarios we have simulated 2M events.

\section{results}
\label{sec:results}

\begin{figure}[tb]
\begin{center}
\includegraphics[angle=0,width=0.45 \textwidth]{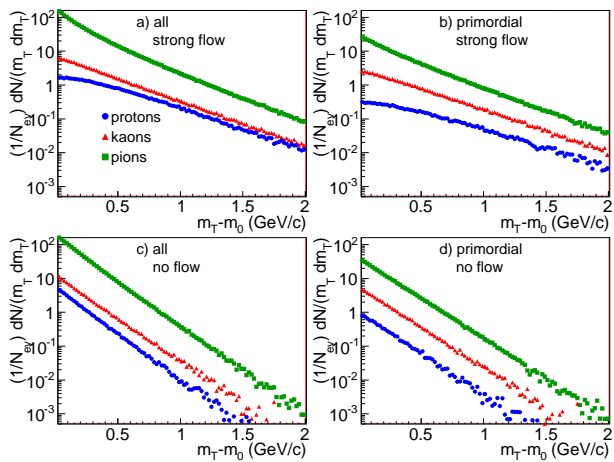}
\end{center}
\vspace{-6.5mm}
\caption{(Color on-line) Single particle inclusive $m_{T}$ spectra for
  pions (green squares), kaons (red triangles) and protons (blue
  circles). Upper panels a) and b) show simulations with ``strong
  flow'', lower panels c) and d) with ``no flow''. Left panels a) and
  c) show all particles, right panels b) and d) only primordial
  particles. 
\label{fig:spectra}}
\end{figure}

Fixing the model parameters, as described above, fixes the average
absolute multiplicity of particles per event. The resulting $\left <
  dN_{ch}/d\eta \right >$ for the ``strong flow'' sample was 8.3. This
is about twice the average multiplicity reported by ALICE
collaboration at $\sqrt{s} = 900$~GeV and is above the average
multiplicity for $\sqrt{s} =
7$~TeV~\cite{Aamodt:2010ft,Aamodt:2010pp,Aamodt:2010jj}.  
 
\begin{table}[tb]
\caption{Slope parameters of the single particle spectra shown in
  Fig.~\ref{fig:spectra}, in MeV. 
\label{tab:temps}}
\begin{tabular}{|r|r|r|r|}
\hline
 & pions & kaons & protons \\
\hline
\multicolumn{4}{|c|}{all}\\
\hline
``strong flow'' & 234 $\pm$ 1 & 323 $\pm$ 4 & 414 $\pm$ 8\\
``no flow''     & 157 $\pm$ 1 & 168 $\pm$ 2 & 157 $\pm$ 3\\
\hline
\multicolumn{4}{|c|}{primordial}\\
\hline
``strong flow'' & 285 $\pm$ 2 & 360 $\pm$ 6 & 465 $\pm$ 2\\
``no flow''     & 183 $\pm$ 1 & 181 $\pm$ 3 & 176 $\pm$ 8\\
\hline
\end{tabular}
\end{table}

\begin{figure}[tb]
\begin{center}
\includegraphics[angle=0,width=0.48 \textwidth]{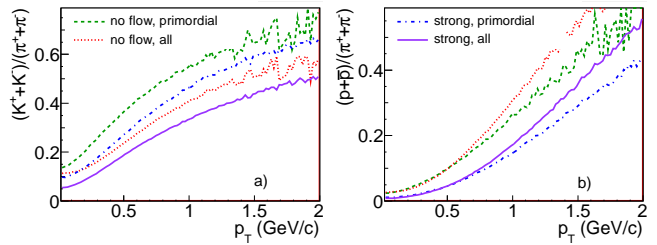}
\end{center}
\vspace{-6.5mm}
\caption{(Color on-line) Kaon to pion (panel a)) and proton to pion
  (panel b)) ratios vs. $p_{\rm T}$. Green dashed lines show primordial
  particles for ``no flow'', red long-dashed show all particles for
  ``no flow'', blue dot-dashed show primordial for ``strong flow'',
  violet full line show all for ``strong flow''.
\label{fig:topiratios}}
\end{figure}

The impact that the radial flow has on particle spectra is shown in
Fig.~\ref{fig:spectra}. Only for the simplest case of
``no flow'' and primordial particles only (shown in panel d)) the spectra
for all particle types are exponential. We fit the distributions
with an exponential and extract their slope, shown in
Tab.~\ref{tab:temps}. Even in the ``no flow'' case the simple
addition of strongly decaying resonances obscures the
picture, as seen in panel c). Resonances tend to populate the lower
$m_T$ regions more abundantly than higher $m_{T}$. There are
relatively more  resonances decaying into pions and protons than into
kaons. As a result, the slope parameter is higher for pions and
protons and lower 
for kaons. For primordial particles in ``strong flow'' in panel b) we
see, as expected, that the thermal ``exponential'' shapes are modified
by radial flow: pions get a ``concave'' shape, while kaons and protons
develop a positive curvature. All are less steep than in the ``no
flow'' case, resulting in the larger slope parameter. Adding resonance
decay products, shown in panel a), decreases the slope parameter by
$50$~MeV, so any attempt to extract the flow velocity and temperature
from the slope 
parameters must fully take into account the resonance contribution. In
Fig.~\ref{fig:topiratios} the kaon to pion and proton to pion ratios
vs. $p_{\rm T}$ are shown. For $K/\pi$ both flow and addition of
resonances decrease the ratio. This comes from the larger contribution
of resonance decays to pions. In contrast, for $p/\pi$ only flow
decreases the ratio, as the relative contributions of resonances to
both pions and protons are similar. In summary, the particle spectra
retain the general characteristics associated with radial flow even
when combined with resonance decays in a small system, but resonance
decays do alter the quantitative estimates of temperature
significantly. In addition our simple model does not include the hard
processes which will contribute to all particle's spectra at larger
$p_{\rm T}$ and complicate the picture further. Therefore making strong
conclusions about radial flow from particle spectra alone is not
trivial and is model dependent.  

Another observable which was extensively used in conjunction with
$p_{\rm T}$ spectra as a signature of collective flow is the pion
femtoscopic radius of the system, more specifically its monotonic
decrease with growing pair momentum $k_{T} = \left | \vec{p_{T,1}} +
  \vec{p_{T,2}} \right |/2$. The first measurement of such a
dependence at the LHC was reported in~\cite{Aamodt:2010jj}. The
minimum-bias measurement shows no significant dependence, however the
combined multiplicity vs. $k_{T}$ analysis shows that the behavior of
``low'' and ``high'' multiplicity events is different. The ``low''
multiplicity (M) events show an increase of pion femtoscopic radius
$R_{inv}$ at low $k_{T}$ and then a fall at high $k_{T}$. The ``high''
M events on the other hand suggest a decrease of the $R_{inv}$ radius
with  $k_T$, although it is also consistent with a flat
dependence. New results from the LHC, especially from 
collisions at the higher energy of $\sqrt{s} = 7$~TeV will deliver
results with better precision and will show if the trend
continues at higher multiplicities.  

We have obtained the pion femtoscopic correlation functions using the
two-particle method described in Section~\ref{sec:blastwave}. We
divide the sample into six $k_{T}$ bins: $(0.12, 0.2)$, $(0.2, 0.3)$,
$(0.3, 0.4)$, $(0.4,0.5)$, $(0.5, 0.6)$ and $(0.6, 0.7)$~GeV/c. We
begin by analyzing the 1-dimensional correlation functions. Following
the experiments~\cite{Aamodt:2010jj,Khachatryan:2010un}, we fit them
with two functional forms, the Gaussian: 
\begin{equation}
f_{G}(q_{inv}) = 1 + \lambda  \exp(-{R^G_{inv}}^2 q_{inv}^2) ,
\label{eq:fitgaus}
\end{equation}
and the exponential:
\begin{equation}
f_{E}(q_{inv}) =1 + \lambda  \exp(-R^{E}_{inv} q_{inv})
\label{eq:fitexp}
\end{equation}
\begin{figure}[tb]
\begin{center}
\includegraphics[angle=0,width=0.4 \textwidth]{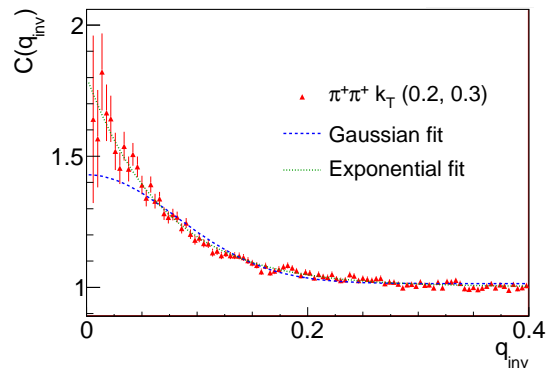}
\end{center}
\vspace{-6.5mm}
\caption{(Color on-line) 1-dimensional pion femtoscopic correlation
  function for the ``strong flow'' case, $k_{T}$ bin of $(0.2,
  0.3)$~GeV/c. Triangles are the results of the simulation, dashed line is a
  Gaussian fit, dotted line is the exponential fit.
\label{fig:fitexample}}
\end{figure}
An example of the fits is shown in Fig.~\ref{fig:fitexample}. We see
a behavior similar to the one reported by the LHC experiments: the
correlation function is better described by an exponential. The
Gaussian fit is worse, but it characterizes the overall width of the
correlation function.  

\begin{figure}[tb]
\begin{center}
\includegraphics[angle=0,width=0.4 \textwidth]{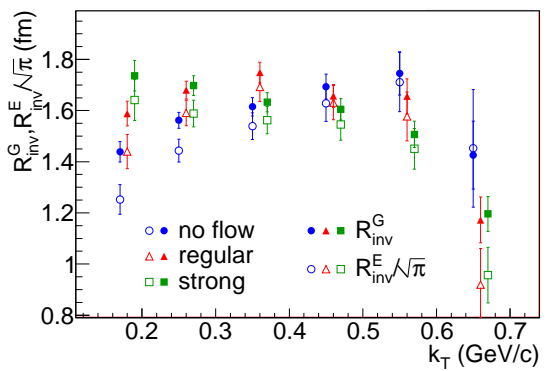}
\end{center} 
\vspace{-6.5mm}
\caption{(Color on-line) Overall 1-dimensional femtoscopic radius
  $R_{inv}$ of the system. ``No   flow'' are shown as blue circles,
  ``regular flow'' as red   triangles,   ``strong flow'' as green 
  squares (the latter two shifted slightly for clarity). Gaussian
  radii $R^G_{inv}$ are shown as closed symbols, exponential radii
  $R^E_{inv}$  as open symbols.
\label{fig:r1ds}} 
\end{figure}
In Fig.~\ref{fig:r1ds} we plot the results of the 1-dimensional
analysis vs. $k_{T}$ for the three scenarios of flow. The Gaussian and
properly scaled exponential fit results are consistent. In the ``no
flow'' scenario the $R^G_{inv}$ is growing with $k_{T}$, similar to low
multiplicity collisions at the LHC. On the other hand, as flow
develops, the $R^G_{inv}$ becomes flat vs. $k_{T}$ for ``regular flow''
and has a negative slope for ``strong flow''. In other words, from the
comparison of the $R^G_{inv}$ vs. $k_{T}$ trends between data and the
model, we observe a development of radial flow with the increase of
multiplicity in the pp collisions at the LHC. It would be interesting
to see if the data at higher multiplicities at $\sqrt{s} = 7$~TeV
continue the trend and develop an even stronger $k_{T}$ dependence.
The results are intriguing enough to attempt a more in-depth
explanation of their origin. We aim to propose more strict tests,
which would make the conclusions from such comparisons less model
dependent.  

\section{3D Bertsch-Pratt analysis}
\label{sec:analysis3d}

\begin{figure}[tb]
\begin{center}
\includegraphics[angle=0,width=0.4 \textwidth]{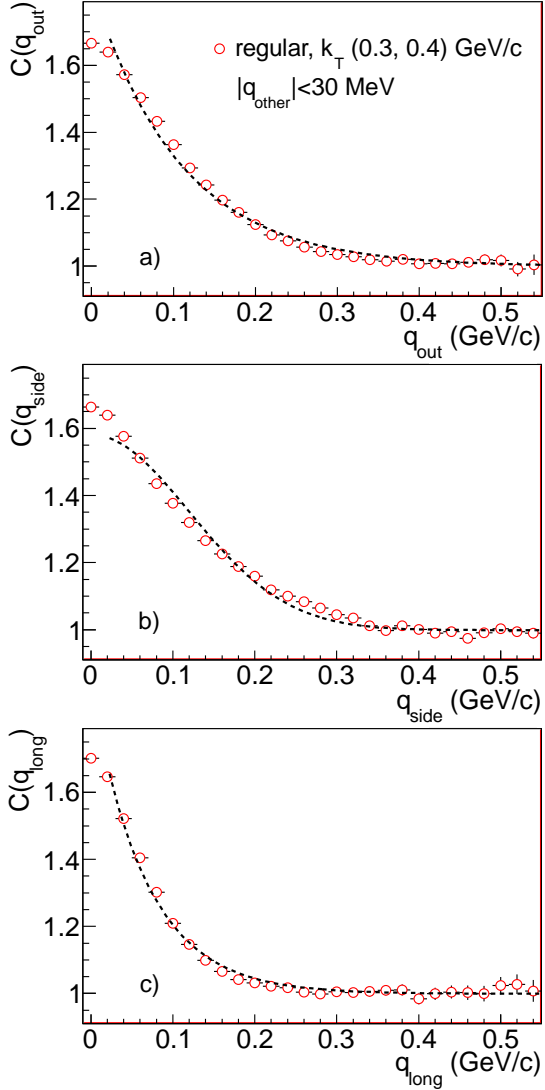}
\end{center}
\vspace{-6.5mm}
\caption{(Color on-line) Projections of the 3-dimensional correlation
  function for the ``strong flow'' simulation for $k_{T} (0.3,
  0.4)$~GeV/c. Dashed lines show fits to 1-d projections. Panel a)
  shows the $out$ projection with an exponential fit. Panel b) shows
  $side$ with Gaussian, panel c) shows $long$ with exponential. To
  plot the projections, other components were integrated in the range
  $|q_{other}|<0.03$~GeV/c. 
\label{fig:3dsources}}
\end{figure}
We have performed the 3D analysis of the simulated correlation
functions. We have used the standard Bertsch-Pratt decomposition of
the relative momentum into the ``long'' (along the beam axis), the
``out'' (along the pair transverse momentum), and the ``side''
(perpendicular to the other two) directions. We have used the
Longitudinally Co-Moving System (LCMS) reference frame, where the 
longitudinal momentum of the pair vanishes. We have analyzed two
classes of pions: the ``primordial'' only, and ``all'' particles. The
latter includes the ``primordial'' plus those from strongly decaying
resonances. In Fig.~\ref{fig:3dsources} we show the correlation
function, separately in $out$, $side$ and $long$ directions. We
attempted to fit both the 1d projections and the full 3d function with
combinations of Gaussian, exponential and Lorentzian shapes. These
three were chosen because they have an analytic relation between
correlation function and pair emission function forms. We found that
(a) for primordial particles a 3D Gaussian is the best fit, (b) for
all particles the $out$ and $long$ direction is best fitted with an
exponential, while $side$ remains Gaussian. This is true both for 1d
projections (shown in the figure) as well as for a full 3d fit. Even
though the fit in Fig.~\ref{fig:3dsources} is best out
of three forms tried, it is still not perfect. Nevertheless, the width
of the function is adequately estimated, which is enough for the
purpose of this paper. The investigation of the details of the shape
of the correlation functions we leave for future studies.
 
Based on the study of correlation function shapes, for primordial
particles a Gaussian 3d ellipsoid, with different sizes in all
directions is a reasonable description of the correlation: 
\begin{equation}
  C_G(\vec{q}) = 1+\lambda \exp(-{R^G_{out}}^2 q_{out}^2-{R^G_{side}}^2
    q_{side}^2-{R^G_{long}}^2 q_{long}^2),
  \label{eq:c3dgaus}
\end{equation}
where $R^{G}_{out}$, $R^{G}_{side}$ and $R^{G}_{long}$ are Gaussian
radii in $out$, $side$ and $long$ directions respectively, $\lambda$
is the strength of the correlation. For all particles we have used the
following fit functional form:  
\begin{equation}
  C_E(\vec{q}) = 1 + \lambda \exp(-R^E_{out} |q_{out}| -
      {R^S_{side}}^2 q_{side}^2 - R^E_{long} |q_{long}|),
    \label{eq:c3dexp}
\end{equation}
where $R^{E}_{out}$ and $R^E_{long}$ are exponential radii in $out$ and $side$
directions. Similarly to the 1-dimensional case, the radii obtained
from Gaussian and exponential fits are not directly comparable, as
they are parameters of different functional forms. 

\begin{figure}[tb]
\begin{center}
\includegraphics[angle=0,width=0.45 \textwidth]{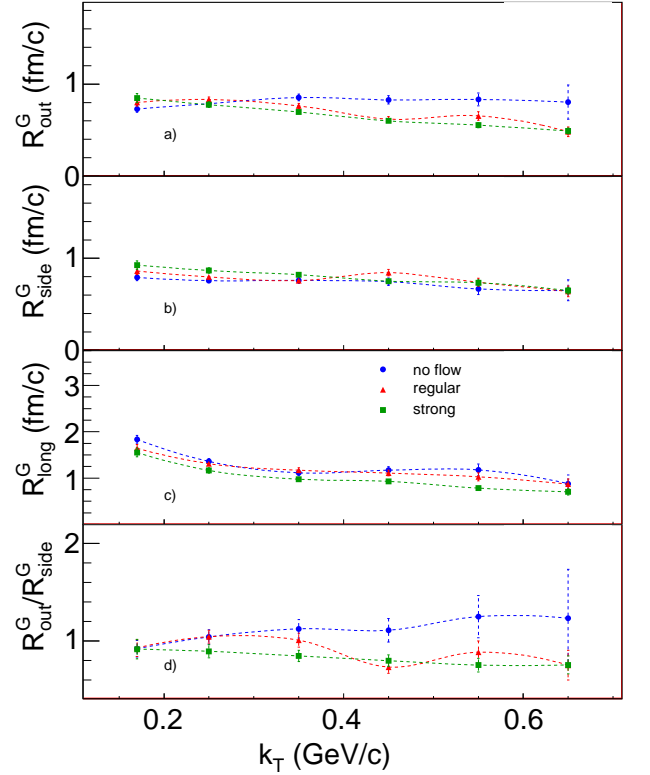}
\end{center}
\vspace{-3.mm}
\caption{(Color on-line) $k_{T}$ dependence of the 3-dimensional radii
  for the primordial particles. Blue circles are ``no flow''
  simulations, red triangles are ``regular flow'', green squares are
  ``strong flow''. Panel a) shows $R^G_{out}$, b) shows
   $R^{G}_{side}$, c) shows  $R^E_{long}$, d) shows
  $R^G_{out}/R^{G}_{side}$ ratio. 
\label{fig:r3dsprim}}
\end{figure}
In Fig~\ref{fig:r3dsprim} we show the Gaussian femtoscopic radii
obtained from the analysis of ``primordial'' particles. We concentrate
on the transverse dynamics. For the $R^G_{out}$ radius we observe an expected
behavior: the ``no flow'' scenario shows a flat $k_{T}$ dependence,
the slope grows with increasing flow and is significant for ``strong
flow''. In addition, the size gets {\it smaller} with increasing flow;
the well-knows mechanism of the ``lengths of homogeneity'' is at work
here. $R^{G}_{side}$ dependence is weakly changed by flow; stronger flow
creates a small slope. The change in behavior is best visible in the
$R^G_{out}/R^{G}_{side}$ ratio. For ``no flow'' it is large, at least
1.0-1.5, for ``strong flow'' it is falling, and smaller, below
1.0. The $R^G_{out}/R^{G}_{side}$ ratio has an advantage that it is,
to a limited degree, independent of scale. The absolute values of the
transverse radii are directly depending on the assumed system size 
(the $\rho_{max}$ parameter), while the ratio is less dependent on it,
and its qualitative behavior should not change as we change
$\rho_{max}$ in a range which would produce multiplicities
observed in pp collisions. 

\begin{figure}[tb]
\begin{center}
\includegraphics[angle=0,width=0.45 \textwidth]{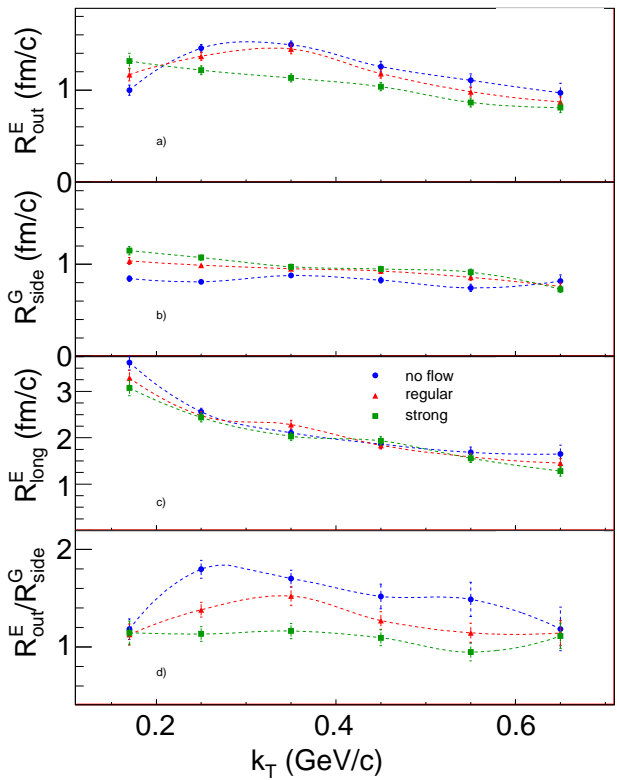}
\end{center}
\vspace{-3.mm}
\caption{(Color on-line) $k_{T}$ dependence of the 3-dimensional radii
  for all particles. Blue circles are ``no flow'' simulations, red
  triangles are ``regular flow'', green squares are ``strong
  flow''. Panel a) show  $R^E_{out}$, b) shows
  $R^G_{side}$, c) shows $R^E_{long}$, d) shows
  $R^E_{out}/R^{G}_{side}$ ratio. 
\label{fig:r3dsall}}
\end{figure}
For primordial particles we have identified the $R^G_{out}/R^{G}_{side}$
ratio as a sensitive, scale-independent, probe of the amount of
collectivity in the system. However, the majority of pions
are coming from resonance decays, which strongly influence spatial and
temporal characteristics of the emission process. It is therefore
crucial to investigate how they modify the signal present for
``primordial'' particles. In Fig.~\ref{fig:r3dsall} we show the fit
results for all pions. We stress that inclusion of resonances changes
the emission patterns visibly, which is reflected in the change of the
fit functional form. $R^E_{out}$ for ``no flow'' is larger than for
``primordial'' and develops a $k_{T}$ slope. In contrast, the 
``strong flow'' $R^E_{out}$ grows little: $R^E_{out}$ grows less as flow
grows. That is surprising, as the ``length of homogeneity'' argument
seems to hold for particles from resonances, even though they do not
come directly from a ``collective'' medium. Also the negative slope of
the $R^E_{out}$ dependence on $k_{T}$ alone is not a good signature of
collectivity as  
such behavior also develops for ``no flow''. $R^{G}_{side}$ shows the
behavior opposite to $R^E_{out}$. For ``no flow'' it is similar to
``primordial'' case, while for ``strong flow'' it grows. As a
consequence, again in striking contrast to the behavior of $R^E_{out}$,
$R^{G}_{side}$ grows more as radial flow increases. Again the
$R^E_{out}/R^{G}_{side}$ ratio is the most sensitive to these changes, and
for all particles it shows an even more dramatic difference between
flow scenarios: the value for the ratio is up to a factor of 1.5
larger for ``no flow'' as compared to ``strong flow''. In summary, we
show that a small value of $R^E_{out}/R^G_{side}$ ratio, close to
unity, is a signal of emission of all particles from a collective
medium. The value is monotonically dependent on the strength of flow:
the stronger the flow, the lower the $R^E_{out}/R^{G}_{side}$ ratio. In
addition, the inclusion of resonance decay products {\it magnifies}
this flow effect.  

\begin{widetext}
\begin{figure*}[tb]
\begin{center}
\includegraphics[angle=0,width=0.9 \textwidth]{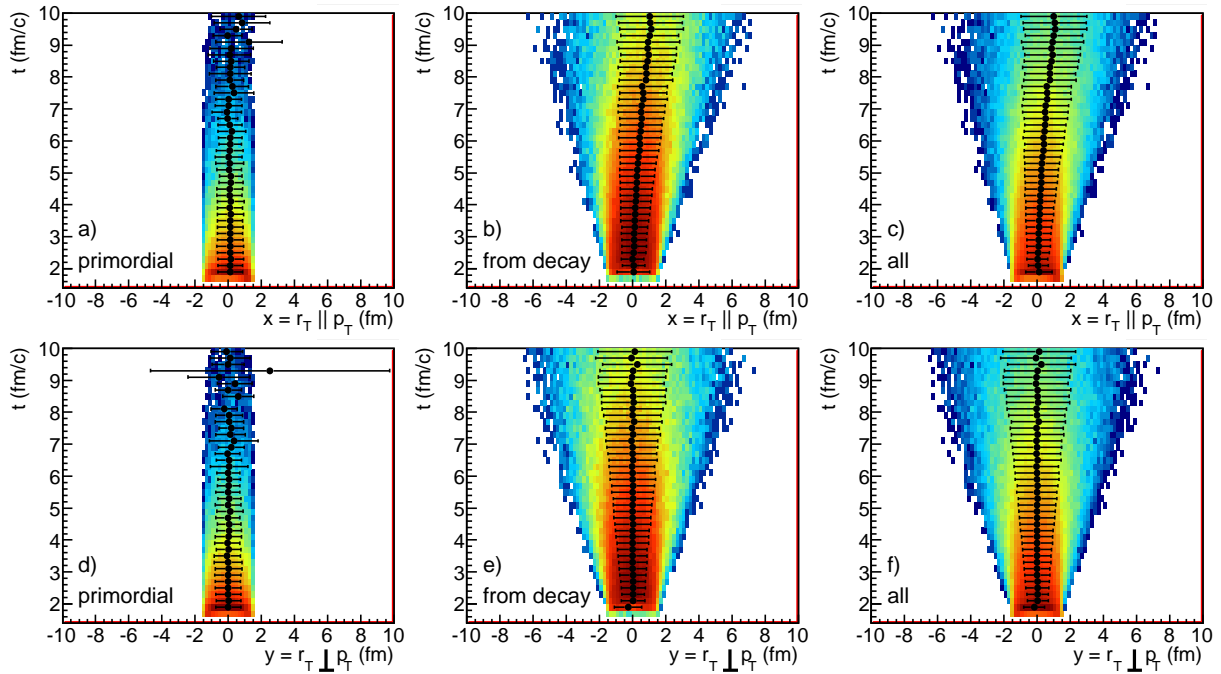}
\end{center}
\vspace{-3.5mm}
\caption{(Color on-line) Correlation between space and time in
  particle emission from the ``no flow'' model. Panels a), b) and c) show
  ``outwards'' direction ($x$ from Eq.~\eqref{eq:routdef}): projection on
  particles' velocity. Panels d), e) and f) show ``sidewards'' ($y$
  from Eq.~\eqref{eq:rsidedef}) direction: projection perpendicular 
  to the velocity. Panels a) and d) show primordial particles, b) and e)
  resonance products, c) and f) all particles. The graphs overlaid
  on the plots show the mean and spread of the distribution in a time bin.
\label{fig:sourcesfl}}
\end{figure*}
 
\begin{figure*}[tb]
\begin{center}
\includegraphics[angle=0,width=0.9 \textwidth]{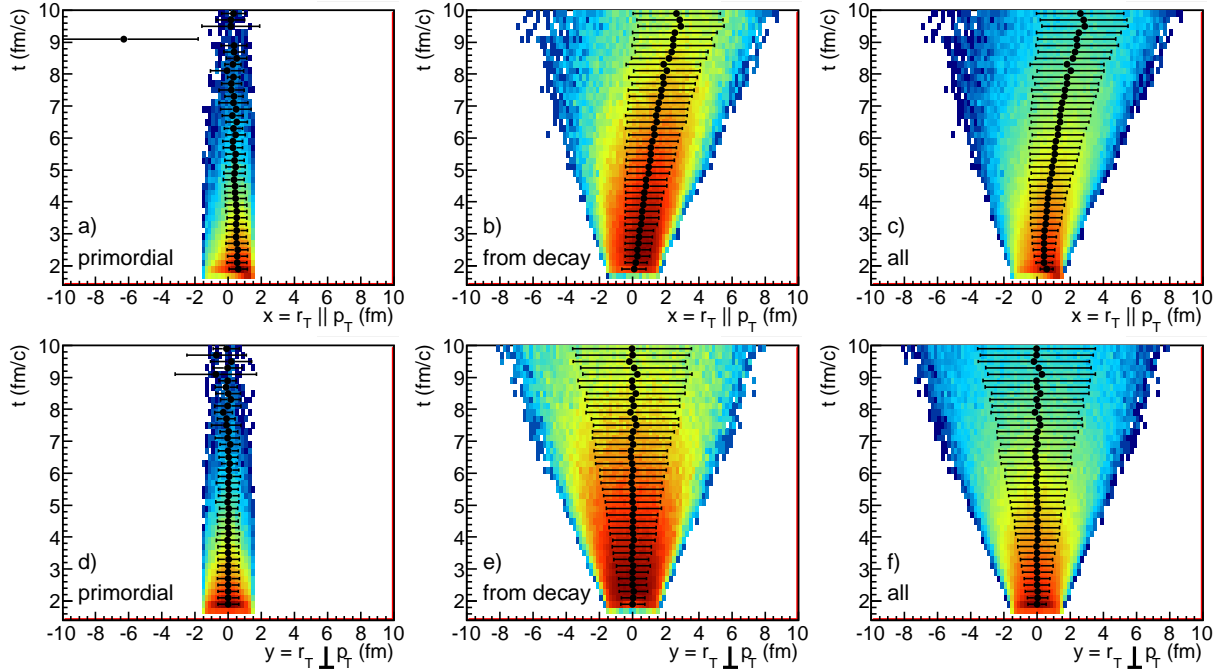}
\end{center}
\vspace{-3.5mm}
\caption{(Color on-line) Correlation between space and time in
  particle emission from the ``strong flow'' model. Panels a), b) and
  c) show ``outwards'' direction: projection on particles'
  velocity. Panels d), e) and f) show ``sidewards'' direction:
  projection perpendicular to the velocity. Panels a) and d) show
  primordial particles, b) and e) resonance products, c) and f) all
  particles. The graphs overlaid on the plots show the mean and
  spread of the distribution in a time bin. 
\label{fig:sourcesmf}}
\end{figure*}
\end{widetext}
 
\section{Origins of femtoscopic radii behavior}
\label{sec:origins}

The identification of the $R^E_{out}/R^{G}_{side}$ ratio as a sensitive
probe of collective particle emission in systems resembling pp
collisions at the LHC has consequences for experiments. It allows to
test a  hypothesis of a creation of a collective system in such
collisions, which produces large particle multiplicities. In addition,
the femtoscopic pion analysis is practically a ``minimum-bias''
measurement and does not require advanced triggering or acceptance
corrections. It can therefore be performed quickly, with the data
samples that the LHC experiments already have on tape. It is therefore
critical to understand why such effect arises and how general their
emergence is; in other words how model dependent is the link between
$R^E_{out}/R^{G}_{side}$ ratio and system collectivity.  

The separation between two particles in the $out$ direction is:
\begin{equation}
r_{out} = x_{1} - x_{2} + \beta(t_2 - t_1)
\label{eq:routdef}
\end{equation}
where $x = \rho p_{\rm T} \cos(\phi-\varphi) /p_{\rm T}$ are the transverse
emission point coordinates of the first and the second particle,
projected on the pair velocity, $t_{1}$ and $t_{2}$ are the emission
times of the particles and $\beta$ is the pair velocity (we neglect
the differences between particles velocities since in order to be
femtoscopically correlated their relative velocity must be small). We
order the particles in a pair in such a way that $t_{2} > t_{1}$, the
third term in the formula reflects the fact that the first particle
has already traveled some distance before the second was born. The
definition of the $side$ direction is that the pair velocity in this
direction vanishes, so the formula for the separation is simpler: 
\begin{equation}
r_{side} = y_{1} - y_{2}
\label{eq:rsidedef}
\end{equation}
where $y = (\rho p_{\rm T} \sin(\phi-\varphi))/p_{\rm T}$ are the transverse
emission point coordinates of the first and second particle, now
perpendicular to the pair velocity. $x$ coordinates are on the x axis
of the top panels of Figs.~\ref{fig:sourcesfl} and~\ref{fig:sourcesmf},
$y$ are on the x axis of the bottom panels and $t$ is on the y axis of
all panels. The $R^{E,G}_{out}$ and $R^{G}_{side}$ femtoscopic radii have the
meaning of variances of $out$ and $side$ distributions respectively
(conventionally divided by $\sqrt{2}$). They are proportional to a
combination of the single particle space-time distribution moments:   
\begin{eqnarray}
R^{E,G}_{out} &\approx& \sigma_{x} + \beta \sigma_{t} - C(x,\beta t) 
\label{eq:Rodef} \\
R^{G}_{side} &\approx& \sigma_{y}
\label{eq:Rsdef}
\end{eqnarray}
where $\sigma_x$, $\sigma_{y}$ and $\sigma_t$ are the variances of the
$x$, $y$ and $t$ distributions respectively and $C(x, \beta t)$ is the
term that accounts for a possible space-time correlation in the
emission process. We note that $C$ contributes to the size with a
negative sign: it means that positive $x-t$ correlation $decreases$
the $out$ radius~\cite{Ko:2006rw, Pratt:2008qv}.

In Fig.~\ref{fig:sourcesfl} the space-time characteristics of the
emission process for the ``no flow'' scenario is shown. On 
panels a) and d) are primordial particles. In this case both out and
side directions show similar size, which does not depend on
time. Also, the mean emission points are not shifted. Panels b) and e)
show the behavior for resonance decay products. There is a 
small correlation between mean $x$ and time. Also, out and side
spreads grow with time, but only slightly. As a consequence, for all
particles shown in panels c) and f), the out emission point is weakly
correlated with time and the side spread grows only slightly.

In Fig.~\ref{fig:sourcesmf} we present a similar study for the
``strong flow'' scenario. We concentrate on the differences between
this scenario and the ``no flow'' one. For primordial
particles, the out mean position is shifted by flow. The spread is
slightly smaller than side; hence the smaller than one ratio for green
points in panel a) of Fig.~\ref{fig:r3dsprim}. For resonance
products we see more striking differences. The out mean position is
strongly positively correlated with time. The spread also grows with
time, but less than for the side, where the growth is more
pronounced than for the ``no flow'' case. As a result, for all
particles, the out mean position is positively correlated with time,
while the side spread grows stronger than in the ``no flow'' case.

We can now understand the dramatic change of the
$R^E_{out}/R^{G}_{side}$ ratio between ``no flow'' and ``strong flow''
scenarios. For ``no flow'', $\sigma_x$ and $\sigma_y$ are similar, as
can be seen in panels c) and f) of Fig.~\ref{fig:sourcesfl}. But
$R^E_{out}$ is further increased by $\sigma_t$ and the $x-t$
correlation is too small to counter-balance it. As a result
$R^E_{out}/R^{G}_{side}$ is large. In short, resonance decays
introduce significant spread in emission times, which acts to increase
$R^E_{out}$. The fact that resonance decay points are distributed
according to the exponential decay time multiplied by velocity also
explains why the shape of the correlation in the $out$ direction
changes dramatically, while it stays Gaussian in $side$, where the
velocity vanishes.

In the ``strong flow'' scenario we note that even for primordial
particles, seen in panel a) and d) of Fig.~\ref{fig:sourcesmf},
$\sigma_x$ is smaller than $\sigma_y$ - the effect of flow of
primordial pions. The effect is further increased by resonances
products. Because of the existence of common radial flow in the system, 
when resonances are born on the freeze-out hypersurface, they are already
flying to the outside of the system. Since they decay after some time,
the emission points of the daughter pions will, on average, be further
apart compared to the ``no flow'' case. This is nicely
illustrated as the strong growth of the $\sigma_y$ with time on panel
e) of Fig.~\ref{fig:sourcesmf}. The result is a significant increase of
$R^{G}_{side}$, as compared to the ``no flow'' case. One expects the
same argument to hold for $\sigma_x$, and to some extent it is seen in
panel b) of Fig.~\ref{fig:sourcesmf}. However, the growth is not as
strong as for $\sigma_y$ - this is because the resonance emission
points are also governed by flow, so their parent $\sigma_x$ is smaller
than $\sigma_y$, just like for primordial pions on panels a) and
d). But more importantly, radial flow of resonances also results in
much stronger (as compared to ``no flow'' case) positive $x-t$
correlation, seen most clearly in panel b). This correlation also
dominates when we consider both primordial and resonance daughter
pions in panel c). This will further decrease the $R^E_{out}$ 
parameter, due to the $C(x,\beta t)$ term in Eq.~\ref{eq:Rodef}. As a
result, the $R^E_{out}/R^{G}_{side}$ ratio is small. In short, radial flow
of primordial pions and resonances produces several effects which all 
combine together to bring the $R^E_{out}/R^{G}_{side}$ ratio down, even
below unity if the flow is strong enough. The stronger the flow, the
lower the ratio.  

We also now understand why 1-dimensional $R_{inv}$ radii dependence on
$k_T$ changes as flow develops. $R_{inv}$ is approximately equal to
the quadratic average of $\gamma R_{out}$, $R_{side}$ and
$R_{long}$, where the $\gamma$ factor in front of $R_{out}$
comes from the boost (with pair velocity) from the LCMS to Pair Rest
Frame, where $R_{inv}$ is determined. For ``no flow'' case $R_{out}$
is flat with $k_T$ in LCMS, so the growth of the boost as $k_T$
increases drives $R_{inv}$ up. For cases with flow, the growth of the
boost is countered by the fall of $R_{out}$ resulting from flow. And
since the other two radii also decrease with $k_T$, the slope of the
$R_{inv}$ dependence on $k_T$ becomes negative.

As a final note we stress that the model employed here
is a simple hydrodynamics-inspired parametrization. We have shown
that one of its particular features - an $\rho-p_{\rm T}$ correlation, also
known as radial flow, produces specific signatures in single
particle spectra as well as 1D and 3D pion femtoscopic radii. However
it does not address the question of where such correlations come
from. One possibility is that they come from a ``medium'' that is
described by hydrodynamic equations, as suggested
in~\cite{Werner:2010ny}, but there may be other mechanisms that
produce such correlations.  

\section{Conclusions}
\label{sec:conclusions}

We have performed simulations with the {\tt THERMINATOR} model, employing
emission of primordial particles from a blast-wave hypersurface, and
including the propagation and decay of a full set of hadronic
resonances. The parameters of the calculations have been adjusted, so
that they produce multiplicities observed in high multiplicity pp
collisions at LHC energies. Three sets of parameters have been
used, which differ only by the strength of the radial flow (the
$\rho-p_{\rm T}$ correlation) in the system.

The radial flow modifies the particle spectra in the expected way,
creating a concave shape for pions and positive curvature for kaons
and protons. However, addition of resonance decays significantly
modifies these shapes, influencing the extracted slope parameter and
making a direct extraction of flow velocity from spectra measurements
alone complicated and model dependent.

Results of 1-dimensional femtoscopic analysis show that with no flow,
the 1D $R^G_{inv}$ pion femtoscopic radius is expected to grow with
pair momentum $k_{T}$. In contrast, as flow is increased, the
dependence changes to a decrease with $k_{T}$. Therefore, within the
frame of this model the results of ALICE collaboration on multiplicity
vs. $k_{T}$ dependence can be interpreted as signature of the 
development of radial flow with increasing per-event particle
multiplicity. 

3D pion femtoscopic radii show that the shape of the correlation
is exponential in the $out$ and $long$ directions due to contributions
from resonance decay products, while in $side$ the shape remains
Gaussian. Radial flow (or lack of it) for primordial pions and
resonances has opposite effect on $R^E_{out}$ and $R^{G}_{side}$
radii. The first grows less as flow develops, which is a result of a
combination of two effects: (a) decrease of lengths of homogeneity for
primordial pions and resonances and (b) strong $x-t$ correlations
coming from resonances' flow-ordered velocities. $R^{G}_{side}$ grows
more as flow develops, the effect  of the preferential propagation of
the resonances to the outside of the source in the presence of
flow. We propose to use the $R^E_{out}/R^G_{side}$ ratio as a probe of
the amount of radial flow (e.g. velocities ordering) in the small,
resonance-dominated system, produced in the high multiplicity
collisions at the LHC. Any analysis aiming to search for collective
effects in such systems should combine at least the three
measurements: identified particle $p_{\rm T}$ spectra, 1D and 3D femtoscopic
radii.    

\section*{Acknowledgements}
The author would like to thank Yiota Foka and Dariusz Mi\'skowiec for
helpful comments and discussions.

\bibliography{ref-rr}

\begin{thebibliography}{23}
\expandafter\ifx\csname natexlab\endcsname\relax\def\natexlab#1{#1}\fi
\expandafter\ifx\csname bibnamefont\endcsname\relax
  \def\bibnamefont#1{#1}\fi
\expandafter\ifx\csname bibfnamefont\endcsname\relax
  \def\bibfnamefont#1{#1}\fi
\expandafter\ifx\csname citenamefont\endcsname\relax
  \def\citenamefont#1{#1}\fi
\expandafter\ifx\csname url\endcsname\relax
  \def\url#1{\texttt{#1}}\fi
\expandafter\ifx\csname urlprefix\endcsname\relax\def\urlprefix{URL }\fi
\providecommand{\bibinfo}[2]{#2}
\providecommand{\eprint}[2][]{\url{#2}}

\bibitem[{\citenamefont{Aamodt et~al.}(2010{\natexlab{a}})}]{Aamodt:2009dt}
\bibinfo{author}{\bibfnamefont{K.}~\bibnamefont{Aamodt}} \bibnamefont{et~al.}
  (\bibinfo{collaboration}{ALICE}), \bibinfo{journal}{Eur. Phys. J.}
  \textbf{\bibinfo{volume}{C65}}, \bibinfo{pages}{111}
  (\bibinfo{year}{2010}{\natexlab{a}}), \eprint{0911.5430}.

\bibitem[{\citenamefont{Aamodt et~al.}(2010{\natexlab{b}})}]{Aamodt:2010ft}
\bibinfo{author}{\bibfnamefont{K.}~\bibnamefont{Aamodt}} \bibnamefont{et~al.}
  (\bibinfo{collaboration}{ALICE}), \bibinfo{journal}{Eur. Phys. J.}
  \textbf{\bibinfo{volume}{C68}}, \bibinfo{pages}{89}
  (\bibinfo{year}{2010}{\natexlab{b}}), \eprint{1004.3034}.

\bibitem[{\citenamefont{Aamodt et~al.}(2010{\natexlab{c}})}]{Aamodt:2010pp}
\bibinfo{author}{\bibfnamefont{K.}~\bibnamefont{Aamodt}} \bibnamefont{et~al.}
  (\bibinfo{collaboration}{ALICE}), \bibinfo{journal}{Eur. Phys. J.}
  \textbf{\bibinfo{volume}{C68}}, \bibinfo{pages}{345}
  (\bibinfo{year}{2010}{\natexlab{c}}), \eprint{1004.3514}.

\bibitem[{\citenamefont{Aamodt et~al.}(2010{\natexlab{d}})}]{Aamodt:2010jj}
\bibinfo{author}{\bibfnamefont{K.}~\bibnamefont{Aamodt}} \bibnamefont{et~al.}
  (\bibinfo{collaboration}{ALICE}) (\bibinfo{year}{2010}{\natexlab{d}}),
  \eprint{1007.0516}.

\bibitem[{\citenamefont{Aamodt et~al.}(2010{\natexlab{e}})}]{Aamodt:2010my}
\bibinfo{author}{\bibfnamefont{K.}~\bibnamefont{Aamodt}} \bibnamefont{et~al.}
  (\bibinfo{collaboration}{ALICE}), \bibinfo{journal}{Phys. Lett.}
  \textbf{\bibinfo{volume}{B693}}, \bibinfo{pages}{53}
  (\bibinfo{year}{2010}{\natexlab{e}}), \eprint{1007.0719}.

\bibitem[{\citenamefont{Khachatryan
  et~al.}(2010{\natexlab{a}})}]{Khachatryan:2010xs}
\bibinfo{author}{\bibfnamefont{V.}~\bibnamefont{Khachatryan}}
  \bibnamefont{et~al.} (\bibinfo{collaboration}{CMS}), \bibinfo{journal}{JHEP}
  \textbf{\bibinfo{volume}{02}}, \bibinfo{pages}{041}
  (\bibinfo{year}{2010}{\natexlab{a}}), \eprint{1002.0621}.

\bibitem[{\citenamefont{Khachatryan
  et~al.}(2010{\natexlab{b}})}]{Khachatryan:2010un}
\bibinfo{author}{\bibfnamefont{V.}~\bibnamefont{Khachatryan}}
  \bibnamefont{et~al.} (\bibinfo{collaboration}{CMS}), \bibinfo{journal}{Phys.
  Rev. Lett.} \textbf{\bibinfo{volume}{105}}, \bibinfo{pages}{032001}
  (\bibinfo{year}{2010}{\natexlab{b}}), \eprint{1005.3294}.

\bibitem[{\citenamefont{Khachatryan
  et~al.}(2010{\natexlab{c}})}]{Khachatryan:2010us}
\bibinfo{author}{\bibfnamefont{V.}~\bibnamefont{Khachatryan}}
  \bibnamefont{et~al.} (\bibinfo{collaboration}{CMS}), \bibinfo{journal}{Phys.
  Rev. Lett.} \textbf{\bibinfo{volume}{105}}, \bibinfo{pages}{022002}
  (\bibinfo{year}{2010}{\natexlab{c}}), \eprint{1005.3299}.

\bibitem[{\citenamefont{Lisa et~al.}(2005)\citenamefont{Lisa, Pratt, Soltz, and
  Wiedemann}}]{Lisa:2005dd}
\bibinfo{author}{\bibfnamefont{M.~A.} \bibnamefont{Lisa}},
  \bibinfo{author}{\bibfnamefont{S.}~\bibnamefont{Pratt}},
  \bibinfo{author}{\bibfnamefont{R.}~\bibnamefont{Soltz}}, \bibnamefont{and}
  \bibinfo{author}{\bibfnamefont{U.}~\bibnamefont{Wiedemann}},
  \bibinfo{journal}{Ann. Rev. Nucl. Part. Sci.} \textbf{\bibinfo{volume}{55}},
  \bibinfo{pages}{357} (\bibinfo{year}{2005}), \eprint{nucl-ex/0505014}.

\bibitem[{\citenamefont{Collaboration}(2010)}]{Collaboration:2010gv}
\bibinfo{author}{\bibfnamefont{C.}~\bibnamefont{Collaboration}},
  \bibinfo{journal}{JHEP} \textbf{\bibinfo{volume}{09}}, \bibinfo{pages}{091}
  (\bibinfo{year}{2010}), \eprint{1009.4122}.

\bibitem[{\citenamefont{Dumitru et~al.}(2010)}]{Dumitru:2010iy}
\bibinfo{author}{\bibfnamefont{A.}~\bibnamefont{Dumitru}} \bibnamefont{et~al.}
  (\bibinfo{year}{2010}), \eprint{1009.5295}.

\bibitem[{\citenamefont{Bozek}(2010)}]{Bozek:2010pb}
\bibinfo{author}{\bibfnamefont{P.}~\bibnamefont{Bozek}} (\bibinfo{year}{2010}),
  \eprint{1010.0405}.

\bibitem[{\citenamefont{Werner et~al.}(2010)\citenamefont{Werner, Karpenko,
  Pierog, Bleicher, and Mikhailov}}]{Werner:2010ny}
\bibinfo{author}{\bibfnamefont{K.}~\bibnamefont{Werner}},
  \bibinfo{author}{\bibfnamefont{I.}~\bibnamefont{Karpenko}},
  \bibinfo{author}{\bibfnamefont{T.}~\bibnamefont{Pierog}},
  \bibinfo{author}{\bibfnamefont{M.}~\bibnamefont{Bleicher}}, \bibnamefont{and}
  \bibinfo{author}{\bibfnamefont{K.}~\bibnamefont{Mikhailov}}
  (\bibinfo{year}{2010}), \eprint{1010.0400}.

\bibitem[{\citenamefont{Humanic}(2007)}]{Humanic:2006ib}
\bibinfo{author}{\bibfnamefont{T.~J.} \bibnamefont{Humanic}},
  \bibinfo{journal}{Phys. Rev.} \textbf{\bibinfo{volume}{C76}},
  \bibinfo{pages}{025205} (\bibinfo{year}{2007}), \eprint{nucl-th/0612098}.

\bibitem[{\citenamefont{Aggarwal et~al.}(2010)}]{Aggarwal:2010bw}
\bibinfo{author}{\bibfnamefont{M.~M.} \bibnamefont{Aggarwal}}
  \bibnamefont{et~al.} (\bibinfo{collaboration}{STAR}) (\bibinfo{year}{2010}),
  \eprint{1004.0925}.

\bibitem[{\citenamefont{Akkelin and Sinyukov}(1995)}]{Akkelin:1995gh}
\bibinfo{author}{\bibfnamefont{S.~V.} \bibnamefont{Akkelin}} \bibnamefont{and}
  \bibinfo{author}{\bibfnamefont{Y.~M.} \bibnamefont{Sinyukov}},
  \bibinfo{journal}{Phys. Lett.} \textbf{\bibinfo{volume}{B356}},
  \bibinfo{pages}{525} (\bibinfo{year}{1995}).

\bibitem[{\citenamefont{Schnedermann et~al.}(1993)\citenamefont{Schnedermann,
  Sollfrank, and Heinz}}]{Schnedermann:1993ws}
\bibinfo{author}{\bibfnamefont{E.}~\bibnamefont{Schnedermann}},
  \bibinfo{author}{\bibfnamefont{J.}~\bibnamefont{Sollfrank}},
  \bibnamefont{and} \bibinfo{author}{\bibfnamefont{U.~W.} \bibnamefont{Heinz}},
  \bibinfo{journal}{Phys. Rev.} \textbf{\bibinfo{volume}{C48}},
  \bibinfo{pages}{2462} (\bibinfo{year}{1993}), \eprint{nucl-th/9307020}.

\bibitem[{\citenamefont{Retiere and Lisa}(2004)}]{Retiere:2003kf}
\bibinfo{author}{\bibfnamefont{F.}~\bibnamefont{Retiere}} \bibnamefont{and}
  \bibinfo{author}{\bibfnamefont{M.~A.} \bibnamefont{Lisa}},
  \bibinfo{journal}{Phys. Rev.} \textbf{\bibinfo{volume}{C70}},
  \bibinfo{pages}{044907} (\bibinfo{year}{2004}), \eprint{nucl-th/0312024}.

\bibitem[{\citenamefont{Kisiel et~al.}(2006{\natexlab{a}})\citenamefont{Kisiel,
  Florkowski, and Broniowski}}]{Kisiel:2006is}
\bibinfo{author}{\bibfnamefont{A.}~\bibnamefont{Kisiel}},
  \bibinfo{author}{\bibfnamefont{W.}~\bibnamefont{Florkowski}},
  \bibnamefont{and}
  \bibinfo{author}{\bibfnamefont{W.}~\bibnamefont{Broniowski}},
  \bibinfo{journal}{Phys. Rev.} \textbf{\bibinfo{volume}{C73}},
  \bibinfo{pages}{064902} (\bibinfo{year}{2006}{\natexlab{a}}),
  \eprint{nucl-th/0602039}.

\bibitem[{\citenamefont{Kisiel et~al.}(2006{\natexlab{b}})\citenamefont{Kisiel,
  Taluc, Broniowski, and Florkowski}}]{Kisiel:2005hn}
\bibinfo{author}{\bibfnamefont{A.}~\bibnamefont{Kisiel}},
  \bibinfo{author}{\bibfnamefont{T.}~\bibnamefont{Taluc}},
  \bibinfo{author}{\bibfnamefont{W.}~\bibnamefont{Broniowski}},
  \bibnamefont{and}
  \bibinfo{author}{\bibfnamefont{W.}~\bibnamefont{Florkowski}},
  \bibinfo{journal}{Comput. Phys. Commun.} \textbf{\bibinfo{volume}{174}},
  \bibinfo{pages}{669} (\bibinfo{year}{2006}{\natexlab{b}}),
  \eprint{nucl-th/0504047}.

\bibitem[{\citenamefont{Yao et~al.}(2006)}]{Yao:2006px}
\bibinfo{author}{\bibfnamefont{W.~M.} \bibnamefont{Yao}} \bibnamefont{et~al.}
  (\bibinfo{collaboration}{Particle Data Group}), \bibinfo{journal}{J. Phys.}
  \textbf{\bibinfo{volume}{G33}}, \bibinfo{pages}{1} (\bibinfo{year}{2006}).

\bibitem[{\citenamefont{Ko}(2006)}]{Ko:2006rw}
\bibinfo{author}{\bibfnamefont{C.~M.} \bibnamefont{Ko}}, \bibinfo{journal}{AIP
  Conf. Proc.} \textbf{\bibinfo{volume}{828}}, \bibinfo{pages}{439}
  (\bibinfo{year}{2006}).

\bibitem[{\citenamefont{Pratt}(2009)}]{Pratt:2008qv}
\bibinfo{author}{\bibfnamefont{S.}~\bibnamefont{Pratt}},
  \bibinfo{journal}{Phys. Rev. Lett.} \textbf{\bibinfo{volume}{102}},
  \bibinfo{pages}{232301} (\bibinfo{year}{2009}), \eprint{0811.3363}.

\end{thebibliography}

\end{document}